\documentclass[usenatbib,useAMS]{mn2e}
\usepackage{epsfig,psfig,graphicx,float,color}
\usepackage{subfigure}
\usepackage{caption}
\usepackage{mybibdefs}
\usepackage{multirow}
\usepackage{widetext}
\usepackage{amsmath}
\usepackage{url}
\usepackage{bm}
\usepackage{float}
\usepackage{color}
\usepackage{amsmath}
\usepackage{mathtools}
\usepackage{listings}
\usepackage{fancybox}


\newcommand{\overbar}[1]{\mkern 1.5mu\overline{\mkern-1.5mu#1\mkern-1.5mu}\mkern 1.5mu}

\DeclareMathOperator*{\argmax}{arg\,max}

\newcommand{\rc}{\textcolor{black}}

\newcommand\ceil[1]{\lceil#1\rceil}

\usepackage{graphicx}
\usepackage{amssymb}
\usepackage{siunitx}
\usepackage{epstopdf}
\pdfoutput=1

\title[Fixing biases from estimating z-distributions]{ Correcting cosmological parameter biases for all redshift surveys induced by estimating and reweighting redshift distributions }
\author[Rau et al.]{Markus Michael Rau$^{1,2}$, Ben Hoyle$^{1}$, Kerstin Paech$^{1,3}$, Stella Seitz$^{1,2}$ \\\\\\\\
$^1$Ludwig-Maximilians-Universit\"at M\"unchen, Universit\"ats-Sternwarte, Scheinerstr. 1, D-81679 Munich, Germany\\
$^2$Max-Planck-Institut f\"ur extraterrestrische Physik, Giessenbachstrasse 1, 85748 Garching, Germany\\
$^3$Excellence Cluster Universe, Bolzmannstr. 2, D-85748 Garching, Germany\\
{\tt E-mail: mmrau@usm.lmu.de}
}

\begin{document}
\date{Accepted ----. Received ----; in original form ----.}
\pagerange{\pageref{firstpage}--\pageref{lastpage}} \pubyear{2014}
\maketitle
\label{firstpage}
\begin{abstract}
Photometric redshift uncertainties are a major source of systematic error for ongoing and future photometric surveys. We study different sources of redshift error caused by \rc{choosing a suboptimal redshift histogram bin width} and propose methods to resolve them. The selection of a too large bin width is shown to oversmooth small scale structure of the radial distribution of galaxies. This systematic error can significantly shift cosmological parameter constraints by up to $6 \, \sigma$ for the dark energy equation of state parameter $w$. Careful selection of bin width can reduce this systematic by a factor of up to 6 as compared with commonly used current binning approaches. We further discuss a generalised resampling method that can correct systematic and statistical errors in cosmological parameter constraints caused by uncertainties in the redshift distribution. This can be achieved without any prior assumptions about the shape of the distribution or the form of the redshift error. Our methodology allows photometric surveys to obtain unbiased cosmological parameter constraints using a minimum number of spectroscopic calibration data. For a DES-like galaxy clustering forecast we obtain unbiased results \rc{with respect to errors caused by suboptimal histogram bin width selection,} using \rc{only 5k} representative spectroscopic calibration objects \rc{per tomographic redshift bin.}
\end{abstract}

\section{INTRODUCTION}
\label{introduction}
Ongoing and future photometric surveys such as DES \citep{DES_ref}, KIDS \citep{2013Msngr.154...44J} and Euclid \citep{Laureijs2011} will photometrically observe hundreds of millions of galaxies. With this rapid increase in statistical power, comes the need to control systematic uncertainties with even higher accuracy, if we wish to remain in the era of precision cosmology. One of the dominant sources of systematic error in these broad band photometric surveys is our ability to obtain accurate distance information characterised by the photometric redshift for the observed galaxies. The accuracy in the photometric redshift distribution for a selected galaxy sample is particularly important, since it enters into the modelling of a wide variety of measurements. Examples of these include projected two point statistics like angular correlation power spectra, or estimates of the critical surface density of a cluster that is important for weak lensing cluster mass measurements \citep{Rau2015, Bonnett2015}. Misestimating the photometric redshift distribution will introduce biases in the respective theoretical models that will \rc{cause errors in the modelling of the signal and hence lead to biased estimates for} e.g. cosmological parameters or cluster masses. 

The main goal of these large area photometric surveys is to improve our understanding of dark energy and the growth of structure. A particularly important probe for this are accurate measurements of two point statistics which are, as mentioned, quite sensitive to errors in the photometric redshift distribution. Since the lack of accuracy in photometric redshift estimates already challenges current multiband photometric surveys like CFHTLens \citep{Choi2015, Kitching2016}, DES \citep{dessanchez, Bonnett2015} or KIDS \citep{2016arXiv160605338H}, it will likely retain its relevance in the next decade where Euclid will probe the Universe to even fainter magnitudes. 

The primary methods to obtain photometric redshift \rc{point} estimates \rc{and estimates of the redshift distribution for individual galaxies} are template fitting methods \citep[e.g.][]{Koo1985, Benitez2000, Bender2001, 2016MNRAS.460.4258L}, empirical methods based on machine learning \citep[e.g.][]{tpz, bonnet, Rau2015} and combinations of the two \citep{Hoyle2015Dataaugment, 2015arXiv150802484S, 2015arXiv151008080S, 2015arXiv151008073S, 2016MNRAS.460.1371B}. Alternatively one can use cross correlations between photometric and spatially overlapping spectroscopic samples to obtain photometric redshift information \citep{Newman2008, Menard2013}. While these cross correlation techniques show great promise and are already applied to photometric data sets \citep{2016arXiv160605338H, 2016MNRAS.460..163R}, the aforementioned established methods remain the state of the art in photometric redshift estimation and are therefore the main focus of this paper. 

Template fitting uses models of the spectral energy distribution (SED) of the different types of galaxies and fits them against the measured photometry to constrain their redshift. As a limited number of broad photometric bands only provides limited information about the SED, the color space spanned by these templates is typically degenerate. This means that the galaxy photometry can be represented by several SED templates \rc{and redshifts}. If the wrong template is fit to the photometry, large photometric redshift errors can occur that can shift cosmological constraints \citep[][]{Hearin2010}. Empirical methods using machine learning have recently became a popular and accurate method for photometric redshift estimation that often outperform contemporary template fitting techniques \citep{dessanchez}. Instead of using theoretical SED templates to model the mapping between photometry and redshift, these methods `learn' it directly from spectroscopic calibration data. This data is taken from the spatially overlapping regions between a spectroscopic and the photometric survey and thus provides both photometric and spectroscopic information. Flexible machine learning \rc{methods} can then use this data to mimic the mapping between photometry and redshift. The result of this fitting process is a model that can provide photometric redshifts for all galaxies in the photometric data set. The process of fitting these flexible models to the color-redshift space of the calibration data can also be supported by extending the available calibration data using artificial galaxies from simulations or SED templates \citep{Hoyle2015Dataaugment}. In this way we can incorporate our understanding of galaxy evolution and the shape of galaxy SEDs into an otherwise completely data driven process. However empirical methods assume, that the calibration data is representative of the true photometric science sample. If \rc{the calibration data is not representative of the full science sample}, the algorithm can produce biased photometric redshifts, since the model is forced to extrapolate into unknown regions of color-magnitude space. 

While empirical methods based on machine learning naturally dependent on representative spectroscopic data, both methods require them to verify their results. These calibration data sets are usually much smaller than the photometric catalogues for which they provide redshift calibration. The main reason for this is the lack of accurate spectroscopic redshift measurements for faint galaxies. Overlapping spectroscopic surveys are not able to completely cover the faint end of the color-magnitude distribution of the photometric survey, because taking spectra at high magnitudes is extremely expensive and requires long exposure times. As a result, the photometric redshifts of significant portions of the faint photometric science sample cannot be verified using accurate spectroscopic redshifts \citep[e.g.][]{Bonnett2015}. Photometric data from these regions is unreliable for usage in cosmological analyses. Thus it needs to be removed \citep{Cunha2014, Bonnett2015, 2016arXiv160605338H} or small samples of spectroscopic redshifts need to be upweighted to obtain a representative validation catalogue \rc{\citep{dessanchez, Rau2015}}. The spectroscopic redshift distributions constructed on these weighted spectroscopic validation catalogues can then be used to test the quality of the photometric redshift distributions. However as these spectroscopic redshift distributions are constructed with a limited number of data that is strongly upweighted in the sparsely populated high redshift tail, they will be noisy and \rc{thus} show a large error. This limits our ability to accurately validate photometric redshift distributions and thus contributes to the total error of the final measurement. 

\rc{We note that this source of error is relevant independent of the method used to generate photometric redshift estimates. Even methods that fit SED templates and do not directly use spectroscopic galaxies during training, also need to be validated on spectroscopic data. The spectroscopic redshift distributions that are constructed during the validation process are then subject to the aforementioned sources of error. This in turn limits the accuracy of photometric redshift validation. }

In this paper we study how this statistical error propagates into cosmological parameter shifts in a DES like galaxy clustering forecast. The goal is to explore how the error in the redshift distribution can be reduced and how the remaining uncertainty can be incorporated into the parameter likelihood. In \S \ref{subsec:bias_on_cosmo} we will show that the selection of a too large histogram bin width can shift cosmological parameter constraints. Subsequently we compare several different binning strategies to reduce this error and provide guidelines for their successful application. The following section \S \ref{subsec:weightingerror} then considers how cosmological parameter constraints are affected by introducing weights to a sample. Some previous work has been done to incorporate errors in the photometric redshift distribution into the parameter likelihood. Most notably the recent work by \citet[][]{Bonnett2015} uses an analytical model for the bias in the tomographic redshift bins and selects a prior on this parameter by comparing several photometric redshift codes. In contrast we study the application of bootstrap techniques to incorporate the uncertainty in photometric redshifts into the parameter likelihood without imposing a specific model. The bootstrap was used in the work from \citet{Cunha2009, dessanchez} and more recently \citet{Bonnett2015, 2016arXiv160605338H} to quantify photometric redshift uncertainty from statistical shot noise. In \S \ref{correcting_bias} we improve upon the bootstrap by studying the `smoothed bootstrap', a modification of this popular resampling method. In addition to the accurate treatment of the statistical shot noise, this new method is also able to correct for systematic shifts in the parameter likelihoods caused by the selection of a too large bin width. In \S \ref{conclusions} we conclude with a discussion on how these methods can be best applied in ongoing and future large area photometric surveys.

\section{DATA}
\label{data}
To mimic the typical shape of photometric redshift distributions, we use the public galaxy mock catalogue from \citet{Jouvel2009}, that resembles the color-redshift space of future imaging surveys like DES or Euclid. We remove data with large magnitude errors $\sigma_{\rm mag} > 0.1$ to produce a catalogue that contains 13k objects \rc{with a median value in the i band magnitude of $21$}. Spectroscopic surveys measure accurate redshifts for their observed galaxies.
In contrast, the photometric redshifts available to imaging surveys have a higher redshift uncertainty and photometric tomographic redshift bins can therefore have broad, non Gaussian, or even mutimodal shapes \citep[e.g.][]{Benjamin2013, dessanchez, Becker2015, Bonnett2015}. 

If the shape of a distribution shows non Gaussian structure, it becomes increasingly hard to estimate this more complicated function with a limited number of samples. In order to obtain the shape of realistic photometric tomographic redshift bins, we need to simulate the way samples of galaxies are selected by their photometric redshift. We therefore first estimate photometric redshift predictions for our sample using the Random Forest \citep{randForestR} algorithm. This method is a popular algorithm for photometric redshift estimation \citep{tpz, dessanchez, Rau2015} and was adopted as one of the standard photometric redshift codes within the DES collaboration.

\rc{Using 5 band photometry in $g$, $r$, $i$, $z$ and $Y$, we} obtain a photometric redshift performance with a mean and scatter on the residuals $\Delta z = z_{phot} - z_{spec}$ of $\left\langle \Delta z \right\rangle = -0.0010$ and $\sigma(\Delta z) = 0.095$. This is comparable with the photometric redshift performance obtained for the DES science verification data as reported in \citet{dessanchez}. 
Using these predictions, we split the sample into 5 photometric tomographic redshift bins, such that each tomographic bin contains approximately the same number of objects. The true redshifts of these \rc{galaxy} samples is then used to estimate the distribution of the tomographic redshift bins. \rc{We define these five tomographic redshift distributions as the true underlying redshift distributions. In the remaining paper we will draw new catalogues of varying sizes from these tomographic redshift distributions. This allows us to compare various estimators for the redshift distribution on these samples. Their accuracy can then be compared with the true underlying redshift distributions. We note that it is therefore important for this analysis to use simulated datasets, as the true underlying redshift distribution has to be known. In real data this underlying truth is never perfectly known. As will be shown in \S \ref{subsec:bias_on_cosmo} biases can persist even in the presence of a very large number of calibration galaxies. In using the true redshift, we} implicitly assume that the algorithm used to produce the photometric redshift distributions does not contribute to further biases in the redshift distribution. This is an optimistic assumption as there is typically a noticable disagreement between codes \citep[e.g.][]{dessanchez, Bonnett2015} that estimate photometric redshift distributions and this will further contribute to the total error budget. Accordingly the total error of the estimated redshift distributions used in the next sections will likely be higher in practise than assumed here. 

\section{METHODOLOGY}
\label{sec:methods}
In the following section we give a brief introduction into density estimation and describe the methods used to select the smoothing scale, i. e. the bin width, in the density estimate. We discuss the different sources of error in density estimates and describe a resampling method to estimate and correct these errors. Finally we describe the Fisher formalism used to propagate the error in the redshift distribution into shifts in the cosmological parameters.
To avoid confusion between the `bias' in redshift distributions and the `bias' in cosmological parameters, we will refer to biases on cosmological parameters as cosmological parameter `shifts'. 
\subsection{Density Estimation}
\label{subsec:dens_estim}
\begin{figure}
   \centering
  \includegraphics[scale=0.4]{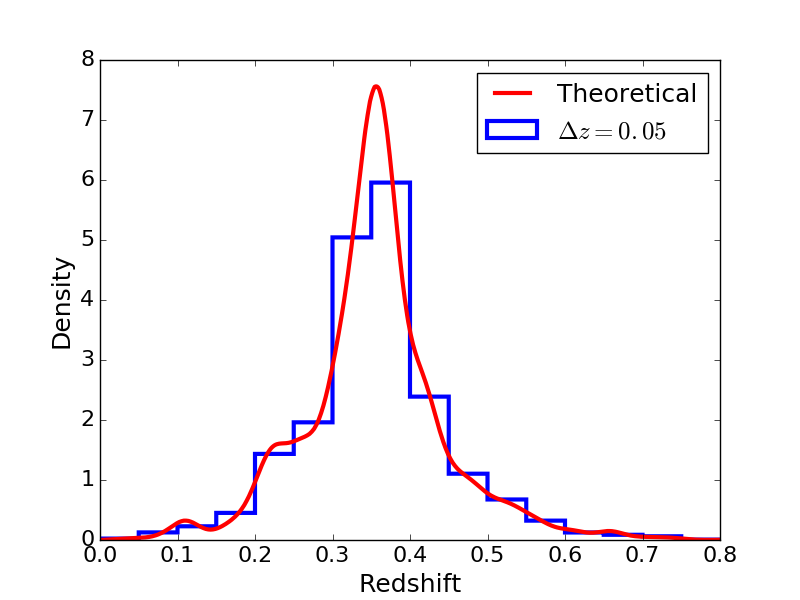}
   \caption{ \label{fig:bias_sampsize} Histogram applied to 16000 objects sampled from the second tomographic bin shown in Fig. \ref{fig:tomographic_distris}. The bin width $\Delta z = 0.05$ was chosen in analogy to \citet{Benjamin2013}.}
\end{figure}

\begin{figure}
   \centering
  \includegraphics[scale=0.4]{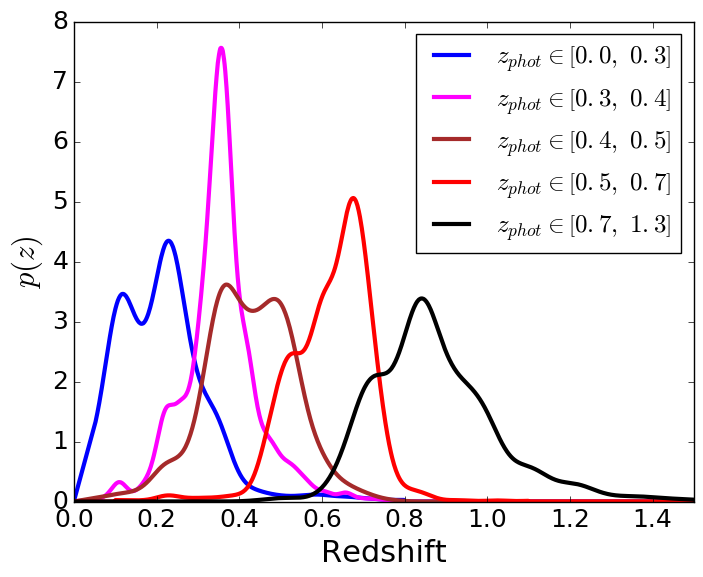}
   \caption{ \label{fig:tomographic_distris} Tomographic redshift distributions generated from the mock catalogue as discussed in \S \ref{data}. The legend shows the photometric redshift bins used to generate the distributions. These tomographic bins are used as reference densities to generate new mock catalogues in a Monte Carlo experiment as explained in \S \ref{cosmo_bias}.}
\end{figure}

The modelling of cosmological observables like angular correlation functions depends on the accurate modelling of the redshift distribution of the tracers. The most common estimator for these distributions is the histogram. To obtain a smooth function that allows for accurate integration, the density at the midpoints is interpolated using spline interpolation. An alternative estimator to obtain smooth density estimates is the kernel density estimate (KDE) that interpolates the density by placing Gaussians on each sample point.\footnote{Instead of Gaussians other so called kernel functions can be used instead. These are non-negative real-valued integrable functions that integrate to unity and exhibit axis symmetry.}  Each of these estimators $\hat{p}(z)$ only approximate the underlying distribution $p(z)$, and their mean squared error can be decomposed into a bias and variance component as
\begin{equation}
E \left[\hat{p}\left( z \right) - p\left(z\right)\right]^2 = {\rm Var}\left\{\hat{p}(z)\right\} + {\rm Bias}^2\left\{\hat{p}(z)\right\} \, , 
\label{eq:MSE}
\end{equation} 
where the variance and bias terms are defined as 
\begin{equation}
{\rm Var} \left\{\hat{p}(z)\right\} = E \left[\left(\hat{p}(z) - E\left[\hat{p}(z)\right]\right)^2 \right] \, 
\label{eq:var_eq}
\end{equation}
and
\begin{equation} 
{\rm Bias}\left\{\hat{p}(z)\right\} = E \left[\hat{p}\left(z \right)\right] - p\left(z\right) \, .
\label{eq:bias_eq}
\end{equation}
\rc{The bias and variance of a density estimate as defined in Eq. \ref{eq:var_eq} and \ref{eq:bias_eq} are functions of redshift and quantify the error from the full shape of the redshift distribution.}
The bias of a density estimate determines how closely the model fits the data. Picking a small bin width leads to a very noisy density estimate. It has a low bias as the small scale features of the particular sample are closely fit. However the variance in this estimate will be large, since some of its bumps may be spurious and not actual features of the underlying distribution to be estimated. Thus the same density estimate, e.g. a histogram with the same fine grained binning, will look quite different for multiple samples independently drawn from the same parent distribution. In contrast, picking a large bin width leads to very smooth functions with a low variance. The density estimate is very stable but can oversmooth important small scale structure in the underlying distribution as shown in Fig. \ref{fig:bias_sampsize}. We randomly sample 16000 galaxies from the second tomographic bin shown in Fig. \ref{fig:tomographic_distris} and apply a histogram with the same bin width used in \citet{Benjamin2013}. The estimate significantly underestimates the peak of the true t-shaped distribution. Increasing the number of samples will not significantly improve upon this result, as the fixed bin width does not allow the histogram to further adapt to the narrow peak. In order to gain improvement, it is necessary to select a smaller bin size, which will also reduce the risk of oversmoothing. As illustrated in this example, the accuracy of a density estimate like the histogram will primarily depend on the amount of smoothing, i.e. the bin- or kernel width, in the estimate\footnote{To simplify the notation, we refer to the bin size as the parameter that governs the smoothing of all density estimates that will be discussed in this work.}. The optimal bin width depends on the number of data samples as well as the shape of the underlying distribution to be estimated. Larger samples allow for smaller bins without producing noisy histograms. Narrow distributions also require narrower bins, to properly estimate the structure of the peak. The selected bin width affects the bias and variance terms in Eq. \ref{eq:MSE} in opposite ways and the tradeoff between both types of error needs to be balanced in any real density estimate to produce the lowest possible total error. 
Thus we need to adapt the bin size as a function of the number of samples and the shape of the distribution. In the following section we will discuss methods that optimize the size of the bins for this purpose.
\subsubsection{The Histogram Estimate}
\label{subsec:histogramm_estim}
The histogram is a density estimate that approximates the underlying density as a step function. The height of each step is proportional to the number of objects falling into a particular grid cell and the smoothing scale of the histogram is determined by the width of these bins. As the modelling of cosmological observables contains integrals over the redshift distribution, software packages like cosmosis \citep{Zuntz2015} interpolate the midpoints of the histograms with splines to perform this integration accurately and efficiently. Following the cosmosis software package we use the Akima spline interpolation scheme \citep{Akima1970}. This method minimizes spurious wiggles for low density values that otherwise pose a problem when using cubic spline interpolation. As the density estimate can still be negative due to numerical errors, we set all negative density values to zero \rc{and renormalize}. 
We want to ensure the same numerical accuracy for all considered density estimates, irrespective of the selected bin width. Thus we evaluate each interpolated histogram on a fixed grid with 1000 equally spaced gridpoints over the whole redshift range $z \in (0.0, 1.5)$. For the following analysis we compare a histogram evaluated on a fixed grid using a bin width of $\Delta z = 0.05$ in analogy to \citet{Benjamin2013}, with a bin width selection scheme that adapts the size of the bin as a function of the number of samples and the shape of the distribution. Assuming a Gaussian reference distribution, one can show \citep{Scott1991} that the optimal bin width for a linearly interpolated histogram is 
\begin{equation}
	h = 2.15 \, \hat{\sigma} \, n^{-1/5} \, ,
	\label{eq:freq_polygon}
\end{equation}
where $\hat{\sigma}$ is the estimator for the standard deviation and $n$ is the number of \rc{galaxies}.\footnote{The bin width relates to the number of bins as $k = \ceil{(\max{z} - \min{z})/h}$.} Even though this rule was derived for the case of a linarly interpolated histogram, it performs well for the case of an Akima spline interpolation as shown later in \S \ref{subsec:bias_on_cosmo}. 
Note that Eq. \ref{eq:freq_polygon} also depends on the shape of the distribution as parametrized by the sample standard deviation. This ensures that the density estimate favours smaller bin widths for strongly peaked distributions which reduces the chance of `oversmoothing' peaks in the density estimate. In the following we will refer to the histogram bin width selection rule defined in Eq. \ref{eq:freq_polygon} as the adaptive bin width selection rule.
\subsubsection{Kernel Density Estimate}
\label{bw_sel_estimators}
The kernel density estimate (KDE) approximates the underlying distribution as a sum of Gaussians centered on the sample points. More formally, the density $p(z)$ is determined by interpolating the density between the weighted sample points $z_i$ with weights $w_i$ using Gaussians $\mathcal{N}\left(z | z_i, \sigma\right)$ centred on each sample point $z_i$
\begin{equation}
\hat{p}(z) = \sum_{i = 1}^{n} w_i \, \mathcal{N}\left(z | z_i, \sigma\right) \, .
 \label{eq:KDE}
\end{equation} 
The standard deviation $\sigma$ of the Gaussians determines how smooth the resulting estimate will be. Broad Gaussians oversmooth the small scale structure of the underlying density and have a similar effect to selecting a too large histogram bin width. In contrast, a standard deviation that is too small can lead to spurious wiggles in the resulting estimate. For simplicity, the following discussion will globally refer to the bin width as the smoothing parameter of kernel density estimates. Assuming that the underlying distribution is Gaussian or close to Gaussian, the Scott rule selects a bin width that minimizes the error in the density estimate. To estimate the underlying density using a sample of size $n$, the optimal value for $\sigma$ is given as
\begin{equation}
  \sigma_{\rm scott} = 1.06 \, \hat{\sigma} \, n^{-1/5} \, ,
  \label{eq:scott}
\end{equation}
where $\hat{\sigma}$ denotes the standard deviation of the sample. 
\subsubsection{The Knuth Rule}
\label{subsub:knuth}
In the previous sections we considered simple rules for bin width selection, that assume parent distributions of close to Gaussian shape. This has computational advantages and also allows for an easy application to weighted data. The next section compares these simple methods with a more advanced method developed in \citet{Knuth2006}, that uses Bayesian inference to fix the number of bins in the histogram. The idea is to maximize the posterior probability for the number of bins $M$ given the data vector $\mathbf{d}$
\begin{equation}
 \hat{M} = \argmax\{\log{p(M | \mathbf{d})}\} \, .
\end{equation} 
Using bayes theorem this posterior is constructed as a nested integral over $\underline{\pi}$ that denotes the vector of probabilities that samples are drawn from each of the $M$ histogram bins
\begin{equation}
p(M | \mathbf{d}) \propto \int {\rm d}\underline{\pi} \, \, p(\underline{\pi} | M) \,  p(M) \,  p(\mathbf{d} | \underline{\pi}, M) \, ,
\end{equation}
where the data likelihood $p(\mathbf{d} | \underline{\pi}, M)$ takes the form of a multinominal distribution, however with a different normalization factor. They continue by choosing a noninformative prior for the bin probabilities $ p(\underline{\pi} | M)$, known as the Jeffreys's prior for the multinominal likelihood \citep{Jeffreys61} and a uniform prior for the number of bins $p(M)$. We use the implementation in the astroML package \citep{astroML} which at the time of this work does not support the application to weighted data. The Knuth rule selects bins of equal width. We also tested the Bayesian Blocks method \citep{Scargle2013} which adapts the width of the individual bins, again using the implementation in the astroML package. However the results we obtained using Bayesian Blocks were much worse compared with all algorithms considered in this work.
\subsubsection{Weighting}
\label{subsub:weighting}
\begin{figure}
   \centering
  \includegraphics[scale=0.4]{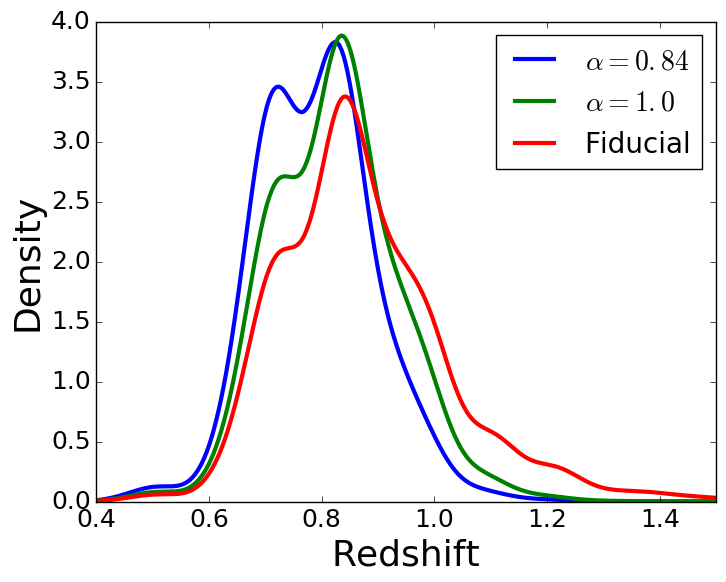}
   \caption{ \label{fig:weighting_distris} Weighting scheme applied to the highest tomographic redshift bin to mimic the lack of spectroscopic calibration data at high redshift. The weighted distributions are generated by multiplying the fiducial analytical distribution of the highest tomographic redshift bin with the sigmoid weighting function defined in Eq. \ref{eq:sigmoid_fnkt}. The distributions are normalized afterwards and the parameter $\alpha$ parametrizes the position of the sigmoid used to penalize the high redshift tail.}
\end{figure}

As already described in the introduction, some weighting schemes are usually applied to the galaxy sample, when computing photometric redshifts. These weights are often the result of empirical photometric redshift codes that interpolate the photometric redshift of a large number of galaxies using a small number of spectroscopic calibration data. The high redshift tail of a photometric redshift distribution is then obtained by giving large weight to a small number of spectrosopic training objects. We mimic this depletion of available spectroscopic objects in high tomographic redshift bins by multiplying a sigmoid weighting function to the highest tomographic redshift bin 
\begin{equation}
	w(z | \alpha) = \left[1 + \exp{\left(10 \left(z - \alpha\right)\right)}\right]^{-1} \, ,
\label{eq:sigmoid_fnkt}
\end{equation}
where $\alpha$ is a parameter that parametrizes the redshift position of the sigmoid. This weighting scheme is illustrated in Fig. \ref{fig:weighting_distris} where the fiducial density (red) is penalized at the high redshift tail by the weighting functions $w(z | \alpha = 1.0)$ (green) and $w(z | \alpha = 0.84)$ (blue). The high redshift tail is supressed after applying the weights which mimics the decrease in the number density of spectroscopically observed galaxies beyond $z > 1.0$. 
During photometric redshift estimation the penalized distributions shown in green and blue would then be remapped onto the red distribution by the introduction of weights. These weights will give more weight to the few galaxies drawn from the high redshift tail of the blue and green curves and downweight the bulk of the distribution at lower redshift. As there are only few objects at high redshift, the histogram constructed on this weighted data set can be quite noisy.
In the analysis in \S \ref{subsec:weightingerror} we study how shifts in cosmological parameters are affected by this increase in noise after applying weights to a sample. This is studied in a Monte Carlo experiment by first generating samples from the penalized distribution with $w(z | \alpha = 0.84)$ (blue line in Fig. \ref{fig:weighting_distris}). For each of the drawn samples we then use the inverse of the weighting function $1/w(z)$ to attach a weight that maps the penalized distribution back onto the original red one. Weighting a sample introduces a correlation between the individual samples, which reduces the statistical power of the full sample. In order to compensate for this, we use the effective sample size \citep{Kish1969}  
\begin{equation}
n_{\rm eff} = \frac{\left(\sum_{i = 1}^{n} w_i \right)^2}{\sum_{i=1}^{n} w_i^2} \, ,
\end{equation}
that replaces $n$ in Eqs. \ref{eq:freq_polygon} and \ref{eq:scott}. 
The mean and the standard deviation of a weighted sample are then computed using equations
\begin{equation}
 \left\langle z \right\rangle = \sum_{i=1}^{n} w_i z_i 
\end{equation}
and 
\begin{equation}
\sigma^2 = \sum_{i=1}^{n} w_i \left( z_i - \left\langle z \right\rangle \right) 
\end{equation}
respectively. 
\subsection{The Smoothed Bootstrap}
\label{sec:bootstrap}
As discussed previously, the total error of density estimates can be split into a bias component and a variance component. The bias quantifies how closely the obtained density estimate fits the underlying distribution. The variance measures the noise in the estimate. Drawing an analogy to a similar cosmological effect, it can be viewed as the shot noise across bins. In practise these contributions need to be estimated using a limited number of samples. This can be done using resampling techniques like the bootstrap. A commonly used way to incorporate the variance in photometric redshift predictions into the analysis is the popular bootstrap method, here `normal bootstrap', as used more recently in \citet{Bonnett2015, 2016arXiv160605338H}. 

We generate new normal bootstrap samples from our available calibration data set by sampling with replacement new data sets of the same size as the original catalogue. In this work we generate 100 of these bootstrap samples and apply the respective density estimate to each of them. Following \citet{Scott1991} the point wise error bands generated by these bootstrapped density estimates reflect the variance of the histogram. 

While the normal bootstrap is able to estimate the variance contribution to the total error, it does not quantify the bias generated by oversmoothing the histogram by picking a too large bin width. 
A resampling method that properly reflects this bias is the smoothed bootstrap \citep{Scott1991}. The basic goal is to estimate both the bias and the variance of a given density estimate. We generate the new smoothed bootstrap data sets by sampling from this density estimate, instead of drawing from the original data set with replacement as done in the normal bootstrap. This smears out the bootstrap samples on the same smoothing scale used to construct the estimate. As for the normal bootstrap, the density estimate is then reapplied to the generated samples and the bias and variance can be measured. This can be seen as a form of Monte Carlo experiment, where our density estimate approximates the true underlying distribution. We reiterate that the only difference between the normal bootstrap and the smoothed bootstrap lies in the way the bootstrap samples are generated. The normal bootstrap generates them by sampling from the data with replacement, the smoothed bootstrap samples from the density estimate whose bias and variance needs to be estimated.

In the following we consider a kernel density estimate (KDE) of the form given by Eq. \ref{eq:KDE} from which it is particularly easy to draw samples. The KDE can be seen as a Gaussian mixture model, where the Gaussians are centered on the sample points. In order to generate a single smoothed bootstrap realisation of a particular density estimate, we first draw $n_{\rm eff}$ samples from the estimate and then reapply the density estimate to this newly generated sample. 
For the aforementioned case of a kernel density estimate constructed on a weighted sample $(w_j, z_j)$ with $j \in \{1, \dots, n\}$ of size $n$, sampling from the density estimate is done as follows:
\begin{enumerate}
	\item Randomly pick one component $j$ with replacement
	\item Draw a sample $z^{*}$ from $\mathcal{N}(z | z_j, \sigma)$
	\item Return $z^{*}$ and the weight $(z^{*}, w_j)$ attributed to $z_j$
\end{enumerate}
In this way we obtain an ensemble of density estimates $\hat{p}^{*}(z)$. If the original density estimate $\hat{p}(z)$ is a proper approximation of the underlying distribution $p(z)$, then samples generated from them should have comparable statistical properties. The bias between the original density estimate $\hat{p}(z)$ and its smoothed bootstrap realisations $\hat{p}^{*}(z)$ should thus be similar to the bias between the true unknown density $p(z)$ and the original density estimate $\hat{p}(z)$. This fact will later allow us to correct for the shifts in the cosmological parameters. The variance of the smoothed bootstrap realizations will also be similar to the variance in the original density estimate. 
Note that the normal bootstrap is the special case of the smoothed bootstrap where $\sigma \rightarrow 0$. In this work we will construct estimates using 100 bootstrap samples.
In the following section, we briefly describe the Fisher forecast formalism \citep[e.g.][]{1998PhRvL..81.2004K, 2002PhRvD..65f3001H, 2009A&A...507..105J} used to propagate the errors in the photometric redshift distribution into shifts in the cosmological parameters. 
\subsection{Forecasting the parameter shifts}
\label{subsec:quant_bias}
In this work, we focus on biases introduced into the modelling of the angular clustering of galaxies, where the corresponding angular correlation power spectrum for a combination of tomographic bins $(i, j)$ is defined as
\begin{equation}
	C_{i, j}(\ell) = \frac{2}{\pi} \int \, W_{i}(\ell, k) \, W_{j}(\ell, k) \, k^2 \, P(k) \, {\rm dk} \, .
\end{equation}
Here $P(k)$ is the matter power spectrum, $k$ is the wavevector and the galaxy clustering window functions for galaxy clustering are defined as
\begin{equation}
	W_{i}(\ell, k) = \int \, b_g(k, z) \, p_{i}(z) \, j_{\ell}[k \chi(z)] \, D(z) \, {\rm dz} \, .
	\label{eq:window_fnkt}
\end{equation}
The modelling of $W_{i}(\ell, k)$ depends on the galaxy-dark matter bias $b_g(k, z)$, the redshift distribution of the galaxy sample $p_{i}(z)$, the comoving distance $\chi(z)$ and the linear growth factor $D(z)$. 

The offset in the angular correlation power spectrum caused by the inaccurate estimation of the redshift distribution is defined as
\begin{equation}
 \mathbf{\Delta_{Cl}} = \mathbf{C(\ell)_{\rm bias}} - \mathbf{C(\ell)_{\rm fid}} \, ,
\end{equation} 
where $\mathbf{C(\ell)_{\rm bias}}$ denotes the vector of angular correlation power spectra estimated using the non optimal estimator for the tomographic redshift distributions and $\mathbf{C(\ell)_{\rm fid}}$ the fiducial, unbiased, vector of angular correlation power spectra obtained from the tomographic distributions in Fig. \ref{fig:tomographic_distris}.

The shift in cosmological parameters $\Delta_{\mathbf{p}}$ caused by this systematic error in the cosmological observable is given to linear order as
\begin{equation}
\Delta_{\mathbf{p}} = \mathbf{F}^{-1} \, \mathbf{D} \, \mathbf{\Sigma}^{-1} \, \mathbf{\Delta_{Cl}} \, ,
\label{eq:para_bias_def}
\end{equation}
where $\mathbf{F}$ is the Fisher matrix defined in Eq. \ref{eq:fisher}, $\mathbf{\Sigma}$ is the corresponding covariance matrix and $\mathbf{D}$ contains the derivatives \rc{with respect to the} cosmological parameters $p_{\alpha}$
\begin{equation}
D_{\alpha, \beta} = \frac{\partial C_{\beta}(\ell)}{\partial p_{\alpha}} \, .
\end{equation}
Here the index $\beta$ runs over all elements in the data vector $\mathbf{C}(\ell)$, i.e. over all auto and cross correlation power spectra.
The Fisher matrix $\mathbf{F}$ is estimated from the data covariance matrix $\mathbf{\Sigma}$ as
\begin{equation}
F_{\alpha, \beta} = \sum_{\ell = \ell_{\rm min}}^{\ell_{\rm max}} \sum_{(i, j), (m, n)} \frac{\partial C_{i,j}(\ell)}{\partial p_{\alpha}} \, \mathbf{\Sigma}^{-1} \, \frac{\partial C(\ell)_{m, n}(\ell)}{\partial p_{\beta}} \, .
\label{eq:fisher}
\end{equation}
The covariance $\mathbf{\Sigma}$ is modelled as
\begin{equation}
	\rc{\Sigma}_{(i, j)}^{(k, l)}(\ell) = A(\ell) \, \left(\overbar{C}^{(i, k)}(\ell) \, \overbar{C}^{(j, l)}(\ell) + \overbar{C}^{(i, l)}(\ell) \, \overbar{C}^{(j, k)}(\ell)\right) \, ,
\end{equation}
where 
\begin{equation}
A(\ell) = \frac{\delta_{\ell, \ell^{'}}}{(2 \ell + 1) f_{\rm sky}} 
\end{equation}
weights the covariance by the inverse fractional sky coverage $f_{\rm sky}$ and $\overbar{C}^{(i, j)}(\ell)$ denotes the angular power spectra estimates, including the shot noise contribution
\begin{equation}
	\overbar{C}^{(i, j)}(\ell) = C^{(i, j)}(\ell) + \frac{\delta_{i, j}}{\overbar{n}_{g}^{i}} \, .
\end{equation}
Here $\overbar{n}_{g}^{i}$ is the number of galaxies per steradian in the respective sample.
The angular correlation power spectra are estimated using the cosmosis software \citep{Zuntz2015}. In this work we follow \citet{2013MNRAS.432.2945H} and \citet{Schafer2015} by using a five dimensional fiducial parameter space $(\Omega_m, w_0, w_a, A_s, n_s)$ in our forecast. This neglects the uncertainty in the parameters $\Omega_b = 0.04$ and $h = 0.72$ which we fix to default values used in the cosmosis software package. \citet{2013MNRAS.432.2945H} and \citet{Schafer2015} justify this simplification by arguing that these remaining parameters are well constrained by other probes like \textit{Planck}. Additionally we marginalize over a multiplicative galaxy-dark matter bias and include modes from $[\ell_{\rm min}, \ell_{\rm max}] = [10, 1000]$ assuming a fractional sky coverage of $f_{\rm sky} = 0.12$ with a number density of $2 \ {\rm arcmin}^{-2}$ for each of the five tomographic redshift bins shown in Fig. \ref{fig:tomographic_distris}. We use the redshift dependent bias model by \citet{Fry1996} 
\begin{equation}
	b_g(z) = 1 + \frac{b_g - 1}{D(z)} \, ,
	\label{eq:fry_bias}
\end{equation}
where $D(z)$ is the linear growth function and $b_g$ is the \rc{F}ry parameter that we set to the fiducial value of $b_g = 1$, such that the fiducial galaxy-dark matter bias model coincides with a constant value $b_g = 1$. 
We summarize the fiducial cosmological parameter values and constraints in Table \ref{tab:fid_paras}.
\begin{table}
\centering
\caption{Fiducial cosmological parameter values and corresponding cosmological constraints for our DES like galaxy clustering forecast.}
\label{tab:fid_paras}
\begin{tabular}{l  l  l}
Parmeter $p$ & Error $\sigma_{p}$ & Fiducial value  \\ \hline \hline
 $\Omega_m$ & 0.013 & 0.3    \\\hline
 $w_0$ & 0.093 & -1.0   \\\hline
 $w_a$ &  0.42 & 0.0 \\\hline
 $A_s$ & $1.6 \cdot 10^{-10}$ & $2.1 \cdot 10^{-9}$  \\\hline
 $n_s$ & 0.023 & 0.96 \\\hline
 $b_g$ & 0.021 & 1.0 \\\hline
\end{tabular}
\end{table}

\section{Cosmological Biases}
\label{cosmo_bias}
A density estimate has two sources of error which contribute to its total mean squared error, as shown in Eq. \ref{eq:MSE}. The bias of the density estimate increases if we introduce more smoothing by choosing a larger bin width. This stabilizes the density estimate but can oversmooth the density thereby destroying its small scale structure. The second contribution to the total error is the variance of the density estimate. This statistical error occurs, since we use a limited number of spectroscopic calibration data to obtain the density estimate, which leads to errors across bins. As a result the same density estimate applied to multiple catalogues independently drawn from the same parent distribution will be different. This effect is larger when the sample size of the catalogue is small. The introduction of weights to the sample can decrease its effective sample size, which further increases the variance of the density estimate. 

In the next \rc{sub}section we will study how errors in the redshift distribution propagate into shifts in the cosmological parameters in a Monte Carlo (MC) experiment. We note that performing a simulation is necessary as the parent distribution from which real spectroscopic data sets are drawn is unkown. A particular catalogue corresponds only to a single realization of the MC experiment. This makes it impossible to estimate the true cosmological parameter shift with respect to this true unknown distribution using real data.

We generate 100 samples from the distributions shown in Fig. \ref{fig:tomographic_distris} and subsequently apply the density estimates discussed in \S \ref{subsec:dens_estim} to these samples. For each method, we end up with an ensemble of 100 density estimates. If the density estimate would be perfect, each of the obtained distributions would coincide with the \rc{(}theoretical\rc{)} parent distributions shown in Fig. \ref{fig:tomographic_distris}. However as we have a limited amount of data available to construct the estimate, it is not possible to obtain perfect estimates of the redshift distribution. \rc{As discussed in \citet{Bonnett2015}, errors on the mean and the width of the redshift distribution are expected to be the dominant source of photometric redshift error for weak gravitational lensing. However it can be expected that higher order statistics like skewness also contribute to the total error budget. This can be especially important for galaxy clustering and cross correlations like galaxy-galaxy lensing. We therefore take a more general approach and include the full shape of the redshift distribution into our analysis without making any assumptions about the shape of the photometric redshift error.} We use the Fisher forecast formalism described in \S \ref{subsec:quant_bias} to estimate the shift in cosmological parameters with respect to the values obtained using the theoretical parent distributions. In order to study the distribution of parameter shifts composed of the 100 Monte Carlo experiments as a function of the number of galaxies available to construct the estimate, we repeat the MC experiment using a variety of different sample sizes. \rc{We investigate sample sizes per tomographic redshift bin ranging from $N_{\rm bin} = 5000$ to $N_{\rm bin} = 30 000$, which results in a fully representative calibration sample of 25k - 150k galaxies distributed in 5 tomographic redshift bins. These numbers have to be compared with the 50k spectroscopic validation objects that are currently used in DES \citep{Bonnett2015} and the 25k used in KIDS \citep[][Tab. 2]{2016arXiv160605338H}. However we also discuss the large sample limit of $N_{\rm bin} = 13 \cdot 10^{6}$ per tomographic redshift bin, which results in $65 \cdot 10^{6}$ for the full sample.}  

In \S \ref{subsec:bias_on_cosmo} we compare the various density estimates using unweighted data, \S \ref{subsec:weightingerror} then investigates the effect of introducing weights. In the following analyses we will normalize the cosmological parameter shifts $\Delta_{p}$ defined in Eq. \ref{eq:para_bias_def} by the fiducial constraint $\sigma_{p}$ for the respective parameter $\mathbf{p}$ quoted in Table \ref{tab:fid_paras}:
\begin{equation}
\mathbf{p}_{\rm norm} =  \mathbf{\Delta_{p}} \big/ \mathbf{\sigma_{p}} \, .
\label{eq:normpara}
\end{equation}
In the following we will refer to the normalized quantities $\mathbf{p}_{\rm norm}$ defined in Eq. \ref{eq:normpara} as the cosmological parameter shifts. 
\subsection{Oversmoothing Errors}
\label{subsec:bias_on_cosmo}
\begin{figure*}
   \centering  
   \includegraphics[scale=0.47]{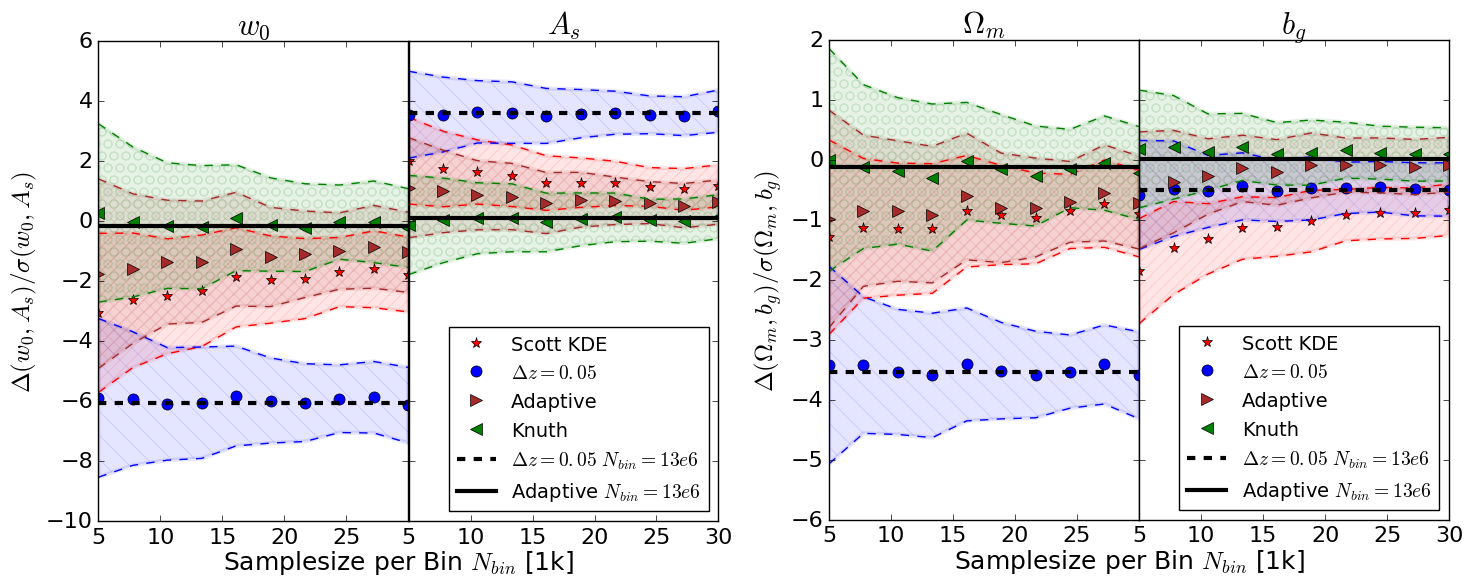}
   \caption{ \label{fig:bias_plot} We show the distribution of cosmological parameter shifts caused by the statistical errors in the estimated redshift distributions as a function of the number $N_{bin}$ of galaxy redshifts available to construct the estimate of the tomographic bin; given in units of 1000 galaxy redshifts. All parameter shifts are normalized by the respective fiducial constraints quoted in \S \ref{subsec:quant_bias}. The left plot considers shifts in the dark energy equation of state parameter $w_0$ and the primordial power spectrum amplitude $A_s$, the right plot in the matter density $\Omega_m$ and the galaxy-dark matter bias parameter $b_g$. We compare the performance of the Scotts rule applied to a kernel density estimate `Scott KDE', a simple histogram with a bin width of $\Delta z = 0.05$, the adaptive bin width selection rule `Adaptive' and the Knuth rule for histograms `Knuth'. The respective distributions of parameter shifts are constructed on 100 simulated catalogues drawn from Fig. \ref{fig:tomographic_distris} as described in the text. The points show the mean \rc{and} the dashed \rc{curves enclose the $\pm 1 \sigma$ error regions} of the respective distribution. The mean of the distribution of relative parameter shifts evaluated on a large sample ($N_{\rm bin} = 13 \cdot 10^{6}$) is illustrated by the horizontal lines. We show this large sample limit for the simple histogram with $\Delta z = 0.05$ \rc{(dashed black line)} and the adaptive bin width selection rule \rc{(solid black line)}.}
\end{figure*}

Fig. \ref{fig:bias_plot} shows the parameter shift distributions in the set of four parameters $(\Omega_m, b_g, w_0, A_s)$ obtained using the different density estimates introduced in \S \ref{subsec:dens_estim}. The results for $w_a$ are not shown, because they are very similar to those obtained for $w_0$ due to the intrinsic correlation between $w_0$ and $w_a$. 
For each parameter we plot the distribution of parameter shifts on the vertical axis as a function of the number of galaxies $N_{bin}$ per tomographic bin in units of 1000 objects on the horizontal axis. We showcase these distributions by the respective mean parameter shift and the standard deviation of parameter shifts. The dashed regions denote the $\pm 1 \sigma$ regions of this distribution, the points denote the mean values. The following discussion refers to the mean and standard deviation of the distribution of parameter shifts as the mean parameter shift and the parameter shift scatter respectively. The horizontal lines illustrate the large sample limit of the respective method, showing the mean parameter shift evaluated on a large sample of $N_{bin} = 13 \cdot 10^{6}$ galaxies per tomographic bin. The parameter shift scatter decreases with increasing sample size for all methods. The relative sensitivity to errors in the redshift distribution strongly depends on the cosmological parameter. While the galaxy-dark matter bias parameter $b_g$ is least sensitive to errors in the redshift distribution, the dark energy equation of state parameter $w_0$ shows large parameter shifts. 
The performance of the four bin width selection algorithms differs especially in terms of their mean parameter shift values. Algorithms that adapt the bin width with the shape of the distribution and the number of objects are consistent, i. e. the mean parameter shift vanishes in the large sample limit. In contrast the histogram with $\Delta z = 0.05$ always produces a large mean parameter shift even in the large sample limit, where the estimator is very stable and the parameter shift scatter vanishes. The histogram with $\Delta z = 0.05$ therefore oversmoothes the underlying distribution. The Knuth rule, being the most sophisticated bin width selection method considered in this work, tightly adapts the histogram to the underlying density and produces very small mean parameter shifts almost independent of the sample size. The parameter shift scatter is however significant for all considered methods even for moderate sample sizes of $N_{bin} = 30000$ per tomographic bin. We reiterate that the parameter shift obtained on a single catalogue will be a single sample from the distribution of parameter shifts, where the parameter shift scatter is its standard deviation. The simple bin width selection algorithms like the Scotts rule for the Kernel density estimate and the adaptive bin width selection rule for the histogram produce larger mean parameter shifts compared with the more elaborate Knuth rule. As the implementations of the Knuth rule currently do not support the application to weighted data, its practical applicability is limited for photometric redshift estimation. 

So far we have considered the performance of density estimators applied to unweighted data. In practice, photometric galaxy samples are typically weighted for the cosmological analysis.
These weights can parametrize the quality of a particular measurement like the error on the measured \rc{galaxy} shape in cosmic shear. Furthermore empirical methods for photometric redshift estimation weight a spectroscopic training sample to resemble a photometric sample. 
In the next section we study how the introduction of weights can affect the distribution of parameter shifts.
\subsection{Weighting Errors}
\label{subsec:weightingerror}
The introduction of weights to a sample introduces an artificial correlation between previously independent samples, which increases the variance of density estimates constructed on the weighted sample. We study the resulting shifts in the cosmological parameters by slightly modifying the experimental setup described in the previous sections. 

Instead of considering an unweighted sample, we generate a weighted sample choosing $\alpha = 0.84$ in Eq. \ref{eq:sigmoid_fnkt}, following the methodology described in \S \ref{subsub:weighting}. The resulting weighted sample then resembles the original distribution of the last tomographic bin. The MC experiment can then be performed as explained in the previous sections, with the only modification that we construct the respective density estimate for the last tomographic bin using a weighted sample. 

We want to study the effect of introducing weights to a sample independently of possible modifications to the error of the density estimate that occurs from changing the bin width. Thus we concentrate on the histogram with a fixed bin width of $\Delta z = 0.05$ instead of adapting the bin width with the shape of the distribution and the effective sample size.
\begin{figure*}
   \centering
  \includegraphics[scale=0.47]{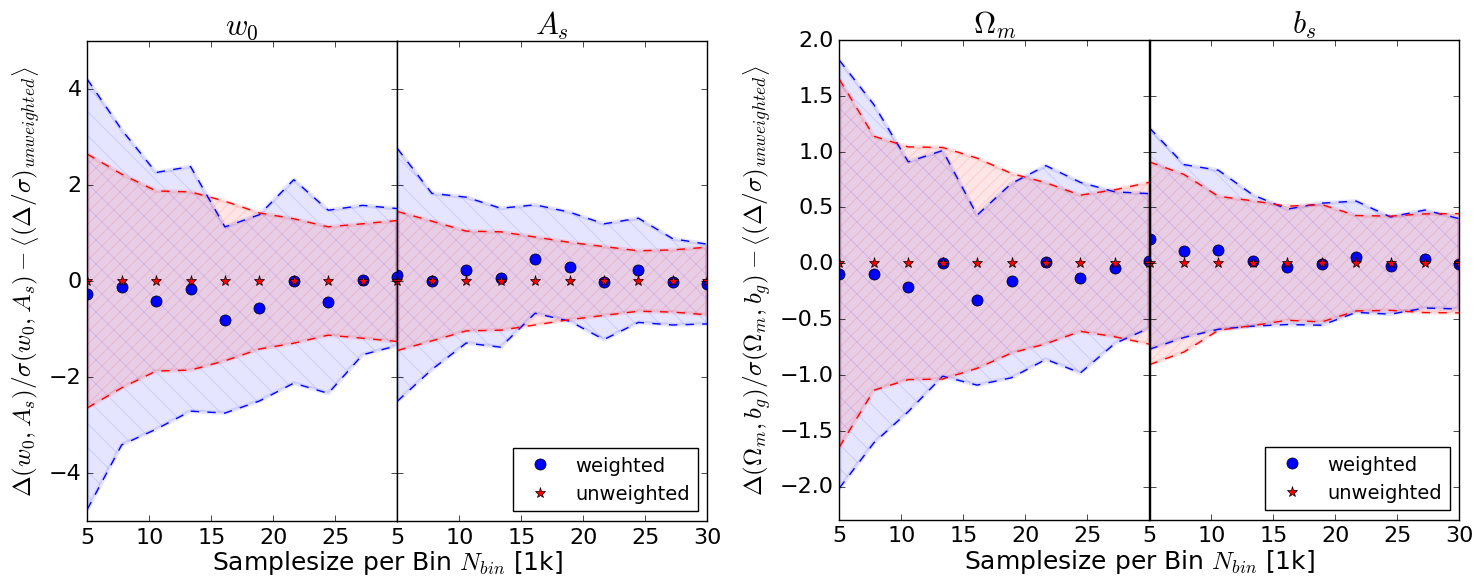}
   \caption{ \label{fig:weighting_om_m_b0} In analogy with Fig. \ref{fig:bias_plot} we compare the distribution of cosmological parameter shifts caused by the statistical errors in the estimated redshift distributions used to reconstruct each tomographic redshift bin for weighted and unweighted data, as a function of the sample size $N_{bin}$. The parameter shifts are normalized by the respective fiducial constraints quoted in \S \ref{subsec:quant_bias}, the sample size $N_{\rm bin}$ is given in units of 1000 galaxies. The dots show the mean, the dashed regions the standard deviation of the distribution of parameter shifts. For both the weighted and the unweighted case, we substract the mean of the distribution of parameter shifts of the unweighted case. Therefore note that the red stars are centered at zero. The left plot considers the dark energy equation of state parameter $w_0$ and the primordial power spectrum amplitude $A_s$, the right plot the matter density $\Omega_m$ and the galaxy-dark matter bias parameter $b_g$. We use the histogram with $\Delta z = 0.05$ as the density estimate.}
\end{figure*}
The result of this experiment is shown in Fig. \ref{fig:weighting_om_m_b0}, where we compare the distribution of parameter shifts for the case of weighted data, with the result for unweighted data. To make the visual comparison between the unweighted and weighted case easier, we substract the mean parameter shift obtained on the unweighted catalogues. In close analogy to the previous section, we show the distribution of parameter shifts for the dark energy equation of state parameter $w_0$, the primordial power spectrum amplitude $A_s$, the matter density  $\Omega_m$ and the galaxy-dark matter bias parameter $b_g$ as a function of the sample size per tomographic redshift bin $N_{bin}$. The parameter shift scatter for the weighted case is in general larger than for the unweighted case. The magnitude of this increase in parameter shift scatter is especially large for the dark energy of state parameter $w_0$ and small for the galaxy-dark matter bias parameter $b_g$. We further note that the mean parameter shifts are only weakly affected compared with the increase in parameter shift scatter. This is to be expected as the introduction of weights to a sample primarily decreases the effective sample size and in turn increases the variance of the density estimate. 

We have seen in the last sections that cosmological parameter constraints can be significantly shifted by errors in the tomographic redshift distributions. The two main sources of error are the effect of oversmoothing and the introduction of weights. Efficient algorithms can closely adapt the bin width to the shape of the distribution and the available sample size to reduce the effect of oversmoothing. However there still remains a statistical error, especially in the presence of weighted samples. In the next section we investigate resampling techniques, that can be used to incorporate both sources of error into the parameter likelihood.

\section{CORRECTING COSMOLOGICAL PARAMETER SHIFTS}
\label{correcting_bias}
\begin{figure}
  \centering
  \includegraphics[scale=0.44]{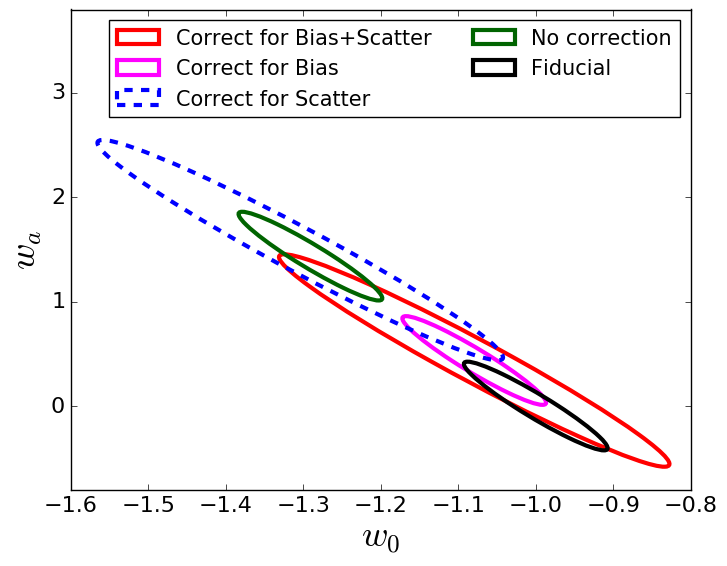}
   \caption{ \label{fig:bias_correction} Typical example of the effect of the cosmological parameter shift correction. The black contour shows the fiducial dark energy parameter constraint. The dark green contour is shifted due to the typical photometric redshift distribution error expected from a KDE with a bandwidth selected by the Scotts rule and applied to a sample of galaxies with $N_{bin} = 5000$. The magenta contour shows the corrected bias using the smoothed bootstrap technique. The red ellipse uses the smoothed bootstrap to marginalize the corrected magenta contour over the remaining statistical uncertainty in the redshift distribution. The dashed blue contour shows the result of marginalizing over the statistical uncertainty using the normal bootstrap technique without the bias correction from the smoothed bootstrap. All contours are $1 \, \sigma$ constraints.}
\end{figure}
In the last section we assumed perfect knowledge of the underlying parent redshift distribution to investigate how the systematic and statistical errors in the redshift histograms lead to shifts in the cosmological parameters. We have seen that the selection of a too large bin width oversmoothes small scale structure in the density estimate. This systematic bias in the redshift distribution propagates into a global shift in the cosmological parameters; the mean parameter shift. This systematic shift is persistent in the large sample limit where the statistical noise in the density estimate vanishes. In addition to this systematic error in the density estimate, we also need to correct for the statistical uncertainty given by the noise in the density estimate. We can incorporate this error into the final parameter constraint by adding its covariance, i.e. the parameter shift scatter, to the fiducial covariance. In practice, the true parent distribution of the tomographic redshift bins is unknown and both sources of error need to be estimated on a single sample. 

This can be done in two steps using the smoothed bootstrap technique as illustrated in Fig. \ref{fig:bias_correction}. In dark green we show a parameter ellipse shifted by the typical error in the redshift distribution obtained from a KDE with a bandwidth selected by the Scotts rule and $N_{bin} = 5000$. \rc{This total sample size of 25k representative spectroscopic calibration objects amounts to approximately the number of spectra used by KIDS \citep{2016arXiv160605338H}.} Using an estimate of the systematic error, i. e. the mean parameter shift, we can correct this biased constraint by shifting it to the magenta contour. Marginalizing over the remaining statistical uncertainty, i. e. the parameter shift scatter\footnote{For simplicity we will refer to estimates of the covariance of the distribution of parameter shifts in two dimensions, e.g. $w_0$ and $w_a$, as the parameter shift scatter, too.}, we can produce the red contour which then almost completely overlaps with the unbiased fiducial contour (black). This has to be compared with the result from the normal bootstrap (dashed blue) that is, like the smoothed bootstrap, able to marginalize over the statistical uncertainty. However in contrast to the smoothed bootstrap, the normal bootstrap is not able to correct for the mean parameter shift. As a result the parameter contour produced by the normal bootstrap is still significantly biased in contrast to the result from the smoothed bootstrap.

\begin{figure*}
   \centering
  \includegraphics[scale=0.45]{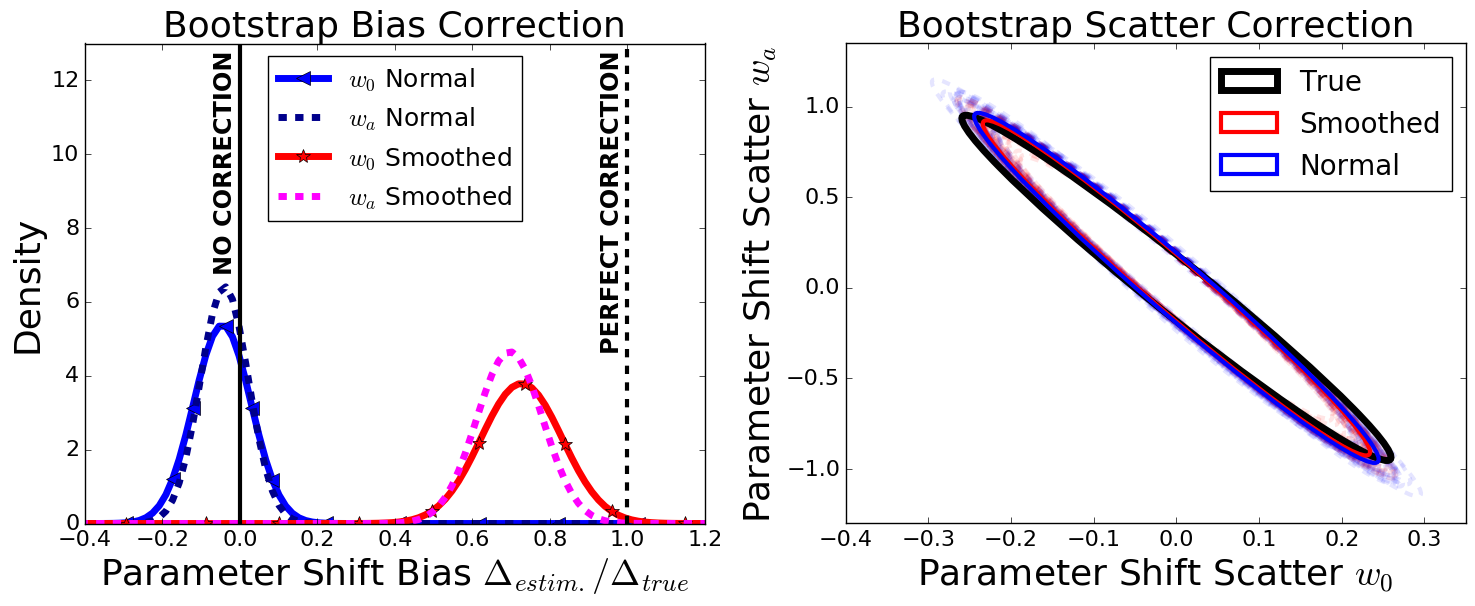}
   \caption{ \label{fig:variance_correction} Left panel: Quality of the cosmological mean parameter shift correction using the smoothed bootstrap compared with the normal bootstrap. The x-axis scale shows the estimated over the true mean parameter shift. The vertical line at the origin corrsponds to no mean parameter shift estimation and therefore no possible correction. The vertical line at 1.0 corresponds to a perfect mean parameter shift estimation and therefore a perfect correction. Right panel: Quality of the variance estimation using bootstrap methods. The black ellipse shows the true $1\sigma$ parameter shift scatter. The red and the meshed red contours show the respective parameter shift scatter estimates from the smoothed bootstrap. The blue and the meshed blue contours show the corresponding result for the normal bootstrap. As explained in the text, the results in both panels are obtained using 50 Monte Carlo experiments, where each time 100 bootstrap samples were drawn.}
\end{figure*}

In the following we compare the smoothed bootstrap technique with the normal bootstrap in a Monte Carlo (MC) experiment. We reiterate, that a simulation is necessary as the true underlying redshift distribution of real samples is unknown. Thus, in order to investigate the statistical performance of the bootstrap techniques, we need to define this true underlying distribution. This experiment is carried out by drawing 50 samples from the theoretical distributions in Fig. \ref{fig:tomographic_distris} and applying the kernel density estimate with a bandwidth selected by the Scotts rule. We choose the kernel density estimate because it is well suited for the generation of new samples which is an important step in the smoothed bootstrap method. To illustrate how we can correct parameter shifts even on a small data set, we choose a sample size of $N_{bin} = 5000$ objects per tomographic bin. For each MC experiment, parameter shifts need to be determined for each of the 100 bootstrap samples. As this is computationally expensive, we choose to perform 50 MC experiments which gives us sufficient statistical accuracy while still being computationally managable. In the following discussion we will refer to the 50 catalogues as the orignal catalogues and to the corresponding 50 density estimates as the original density estimates. For each of the original catalogues we estimate the mean parameter shift and the parameter shift scatter using the smoothed bootstrap and the normal boostrap. We apply the Fisher forecast method to propagate the uncertainties in the redshift distribution into shifts in cosmological parameters in analogy to the previous sections. This gives us 50 estimates for the mean parameter shift and the parameter shift scatter from both resampling methods. The distribution of these estimates is then compared with the true mean parameter shift and scatter obtained on the original density estimates in Fig. \ref{fig:variance_correction}.
The left panel of this figure shows the quality of the estimation of the mean parameter shift using the smoothed bootstrap and the normal bootstrap for the example of the dark energy equation of state parameter. As the mean parameter shift is a constant offset, each estimate should give the same value, independent on the particular sample. In practise this is of course not the case, which we quantify by calculating the mean and the standard deviation of this distribution. For illustration we normalize the respective mean parameter shift estimates by the true mean parameter shift. On this x-axis scale, \rc{0} corresponds to no mean parameter shift correction and 1 corresponds to a perfect correction. 

We find that the normal bootstrap is not able to estimate the mean parameter shift, while the smoothed bootstrap is able to recover the majority of the mean parameter shift. The right panel of Fig. \ref{fig:variance_correction} shows the average quality of the parameter shift scatter estimation for the example of the covariance between $w_0$ and $w_a$. The smoothed bootstrap and the normal bootstrap produce the same estimation quality and are both able to accurately estimate the true parameter shift scatter. The dashed ellipses show the individual parameter shift scatter estimates from the 50 experiments. As can be seen, the scatter around the mean value is in general quite small for both methods.

We have shown that the smoothed bootstrap is able to correct shifts in cosmological parameter constraints produced by errors in the redshift distribution. This includes errors suffered from oversmoothing small scale features in the redshift distribution, as well as statistical errors produced by noisy density estimates.

\rc{The mean parameter shift correction using the smoothed boostrap technique implicitly assumes that the bias between the true unknown distribution $p(z)$ and the original density estimate $\hat{p}(z)$ is approximately equal to the bias between the original density estimate $\hat{p}(z)$ and the density estimates constructed on its smoothed bootstrap samples $\hat{p}^{*}(z)$. In the context of this section, this implicitly assumes that the parameter shift bias is linear with respect to the smoothing parameter. This will likely not be the case in practise. Smoother density estimates are less sensitive to the smoothing parameter, than noisy density estimates, especially in the variance component (Eq. \ref{eq:var_eq}). The density estimates constructed on the smoothed bootstrap samples are smoother than the original estimate. As can be seen in Fig. \ref{fig:bias_plot}, the original density estimate constructed using the Scott rule produces a small mean parameter shift. It is therefore a bit too smooth compared to the true unknown density estimate. As discussed, smoother density estimates will be more stable with respect to changes in the bin width compared with more noisy density estimates. Thus the mean parameter shift between the original density estimate constructed using the Scott rule and the unknown true density estimate will be higher than between the original density estimate and the smoothed bootstrap samples. We identify this non linear dependency of the mean parameter shift on the bin width as the reason why the smoothed bootstrap underestimates the mean parameter shift.}
 
\rc{In contrast the parameter shift scatter is relatively insensitive to changes in the bin width. This can be seen in the Fig. \ref{fig:bias_plot} where the parameter shift scatter values for different bin width selection algorithms have been shown to be quite similar. Thus making the density estimate smoother in the smoothed bootstrap method has an negligible effect on the quality of the estimated parameter shift scatter as shown in Fig. \ref{fig:variance_correction}. The likely reason for this is that the sample size often dominates the error of a statistical estimator over changes in the shape of the distribution. For instance the error on the sample mean scales with $\propto \hat{\sigma} \big/ \sqrt{n}$ where $n$ is the sample size and $\hat{\sigma}$ is the sample estimator for the standard deviation. A small increase in $\hat{\sigma}$ produced by e. g. the smoothed bootstrap is therefore strongly supressed ($1\big/ \sqrt{N_{\rm bin}} = 0.01$) even for a relatively small sample size of $N_{\rm bin} = 5000$ galaxies per tomographic redshift bin as considered here. } 

\section{SUMMARY AND CONCLUSIONS}
\label{conclusions}
\rc{Current and next} generation large area photometric surveys like DES or Euclid are expected to measure cosmological parameters with unprecedented accuracy. To enter this era of precision cosmology, our understanding of systematic errors needs to increase faster than the statistical power of these measurements. Errors in the distribution of distance, or redshift are already challenging for current multiband photometric suveys like CFHTLens, DES or KIDS and are likely to become an even greater burden for next generation surveys like Euclid. 

To prepare for these upcoming challenges, this work studies how photometric redshift distributions can be estimated without causing systematic errors in the cosmological parameters. We start the discussion in \S \ref{subsec:bias_on_cosmo} by considering the statistical properties of a simple histogram estimate of the redshift distribution. We have seen that the selection of a bin width that is too large can bias redshift distributions in each tomographic bin. This `oversmoothing effect' destroys information about small scale features in the density like multimodal or sharp peaks. As a result, the estimated density then no longer coincides with the true underlying distribution. This effect can even be significant for bin widths used in current analysis like the constant redshift binning of $\Delta z = 0.05$ used in \citet{Benjamin2013}. We note that their continued usage can become a significant error source when parameter constraints become tighter in future surveys. 

To reduce these errors in the redshift distribution, we studied methods to adaptively select the bin width as a function of the number of objects and the shape of the distribution. \rc{We specifically investigate small sample sizes of 25k spectroscopic validation galaxies which are currently available in state-of-the-art photometric surveys, where KIDS and DES use approximately 25k and 50k spectroscopic validation galaxies respectively (\citealt[][Tab. 2]{2016arXiv160605338H} and \citealt{Bonnett2015}).} In \S \ref{subsec:bias_on_cosmo} we demonstrated for the case of a DES like galaxy clustering forecast, that these methods can reduce the relative shift in cosmological parameters by a factor of up to 6 for the dark energy equation of state parameters $w_0$ and $w_a$, as compared with the aforementioned constant redshift binnning. We obtained the most accurate redshift distributions using the Knuth rule. Using this method we were able to produce cosmological parameter constraints with an especially low systematic error, even for small sample sizes of 5000 galaxies per tomographic redshift bin.  However current implementations of the Knuth rule do not support the application to weighted data. This severely limits its practical applicability, as some form of weighting scheme is usually applied to redshift samples. The generalization of the algorithm to weighted data should be a straightforward modification of the multinominal data likelihood and the prior on the bin probabilities (see \S \ref{subsub:knuth}). We leave this for future work. Irrespective of the chosen method, there still remains a statistical uncertainty in the redshift distribution that cannot be removed even if the bin width is carefully selected as discussed in \S \ref{cosmo_bias}. 

In \S \ref{subsec:weightingerror} we demonstrated that the introduction of sample weights drastically deteriorates the quality of the redshift histograms. The size of this effect naturally depends on the weighting scheme and the shape of the considered distributions. For our choices the statistical error in the measured dark energy equation of state parameters $w_0$ and $w_a$ increased by a factor of up to two while other parameters like the matter density $\Omega_m$ or the galaxy-dark matter bias $b_g$ were shown to be much more robust.  

The magnitude of the aforementioned types of error scale with the size of the spectroscopic catalogue. In practise the amount of calibration data available for photometric redshift estimation is limited. Especially broad band photometric surveys require accurate calibration as their photometry often does not allow a unique and accurate evaluation of distance, independent of the chosen photometric redshift estimation algorithm.
As spectroscopic surveys use different strategies to select their targets than their photometric counterparts, their selection functions in color-magnitude space are typically different. In particular the estimation of spectra for fainter objects requires long exposure times. Therefore faint regions of color-magnitude space are typically incompletely covered by spectroscopy. 

In order to validate photometric redshifts using spectroscopic surveys, weights need to be introduced such that these incompatible selection functions are corrected \citep[e.g.][]{Cunha2014, Bonnett2015}. In regions of color-magnitude space where no spectroscopic calibration is available, we even need to exclude subsets of the photometric sample to guarantee unbiased results. While the weighted spectroscopic calibration data will mimic the photometric science sample, the resulting density estimates will be more noisy, as the sparsely populated high redshift tail will be strongly upweighted (see \S \ref{subsub:weighting}). This uncertainty in the redshift distribution has to be incorporated into the final parameter likelihood, as it cannot be avoided even by the most accurate bin width selection methods. In \S \ref{correcting_bias} we compared two resampling methods to accomplish this. The first is the commonly used bootstrap method, that failed to correct for the effect of oversmoothing. Instead we showcase a modified version of the bootstrap. The `smoothed bootstrap' smears out the individual bootstrap samples in the same smoothing scale as used in the original density estimate. We demonstrate that this method is able to correct for the effect of oversmoothing to good accuracy. At the same time the smoothed bootstrap shows the same quality in estimating the statistical noise in the redshift distribution as the normal bootstrap. This means that we can accurately marginalize over this remaining statistical uncertainty after the systematic bias from oversmoothing is accounted for. In this way we can correct the final parameter likelihood from both sources of error. 

\rc{While this work mainly adresses redshift distributions in photometric surveys, the results presented here are readily applicable to all problem settings where a distribution needs to be estimated. This includes for example the spectroscopic validation of photometric redshift algorithms that otherwise do not require a representative spectroscopic calibration dataset like template fitting or redshift estimation using cross correlations. In order to accurately validate photometric redshift distributions, we require a sample of representative spectroscopic galaxies. The photometric redshift distribution that has been estimated by any photometric redshift method can then be compared against an estimate of this reference distribution. This distribution of spectroscopic calibration redshifts is however subject to the sources of error discussed in this work. This limits our ability to calibrate photometric redshift distributions which indirectly contributes to the total photometric redshift error of the respective method. We want to highlight that this applies to photometric redshift methods that reconstruct redshift distributions for individual galaxies as well as special photometric redshift point estimates \citep[see e.g.][]{Rau2015} that estimate, in analogy to this paper, redshift distributions of samples of galaxies. It is however especially important for empirical photometric redshift methods based on machine learning. These methods estimate individual object redshift distributions and point predictions by reweighting accurate calibration data in color space. Estimates of the individual object redshift distributions are constructed as weighted density estimates of spectroscopic calibration data. Photometric redshift point predictions can be seen as the mean or median estimated on the weighted calibration dataset. We refer the interested reader to \citep{Rau2015} for a more detailed explanation. As photometric redshift distributions estimated using machine learning are in essence density estimates constructed on a weighted spectroscopic calibration dataset, the methods discussed in this work readily apply to them. The estimation of these weights requires a density estimate in color space that can be the source of additional errors that haven't been explicitly discussed in this work. However the same resampling techniques should also apply here, which we highlight as a direction for future research. While we focussed on the application to the modelling of angular correlation power spectra, we note that the methods developed in this work will also be potentially relevant for other two point statistics like cosmic shear spectra.}

In summary, uncertainties in the photometric redshift distribution are a limiting source of systematic error for ongoing and future photometric surveys. Their quality can only be guaranteed by validating against highly accurate spectroscopic redshift measurements. Weighting methods are able to correct for the mismatch between the spectroscopic and photometric selection functions and the efficient bin width selection algorithms investigated in this work are able to avoid being systematically biased by oversmoothing the resulting density estimates. We finally demonstrated, that the smoothed bootstrap can correct the remaining cosmological parameter biases without assuming a particular model for the redshift uncertainty. In this way future photometric surveys will be able to obtain unbiased cosmological parameter estimates using a minimum amount of spectroscopic calibration data.

\section{ACKNOWLEDGEMENTS}
\label{acknowledgements}
\rc{MMRau thanks the anonymous referee, Gary Bernstein, Michael Troxel and Sarah Bridle for invaluable comments on the manuscript.}
This work was supported by SFB-Transregio 33 `The Dark Universe' by the Deutsche Forschungsgemeinschaft (DFG) and the DFG cluster of excellence `Origin and Structure of the Universe'.

\newpage
\bibliographystyle{mn2e}
\bibliography{bibliography}

\end{document}